\documentclass[pre,twocolumn,superscriptaddress]{revtex4}

\usepackage{amsmath}
\usepackage{graphicx}
\usepackage[colorlinks=true, allcolors=blue]{hyperref}
\usepackage{bm}

\begin{document}

\title{Receptor crosstalk improves concentration sensing of multiple ligands}
\author{Mart\'{i}n Carballo-Pacheco}
\thanks{These authors are listed alphabetically}
\affiliation{School of Physics and Astronomy, University of Edinburgh, Edinburgh, EH9 3FD, United Kingdom}
\author{Jonathan Desponds}
\thanks{These authors are listed alphabetically}
\affiliation{Department of Physics, University of California San Diego, La Jolla, CA 92093, USA}
\author{Tatyana Gavrilchenko}
\thanks{These authors are listed alphabetically}
\affiliation{Department of Physics and Astronomy, University of Pennsylvania, Philadelphia, PA 19104, USA}
\author{Andreas Mayer}
\thanks{These authors are listed alphabetically}
\affiliation{Lewis-Sigler Institute for Integrative Genomics, Princeton University, Princeton, NJ 08544, USA}
\author{Roshan Prizak}
\thanks{These authors are listed alphabetically}
\affiliation{Institute of Science and Technology Austria, Am Campus 1, A-3400, Klosterneuburg, Austria}
\author{Gautam Reddy}
\thanks{These authors are listed alphabetically}
\affiliation{Department of Physics, University of California San Diego, La Jolla, CA 92093, USA}
\author{Ilya Nemenman}
\thanks{Corresponding authors}
\affiliation{Department of Physics, Department of Biology, and Initiative in Theory and Modeling of Living Systems, Emory University, Atlanta, GA 30322, USA}
\author{Thierry Mora}
\thanks{Corresponding authors}
\affiliation{Laboratoire de physique statistique, CNRS, Sorbonne University, University Paris-Diderot and \'Ecole normale sup\'erieure (PSL University), 75005 Paris, France}

\begin{abstract}
Cells need to reliably sense external ligand concentrations to achieve various biological functions such as chemotaxis or signaling.
The molecular recognition of ligands by surface receptors is degenerate in many systems leading to crosstalk between different receptors. 
Crosstalk is often thought of as a deviation from optimal specific recognition, as the binding of non-cognate ligands can interfere with the detection of the receptor's cognate ligand, possibly leading to a false triggering of a downstream signaling pathway. 
Here we quantify the optimal precision of sensing the concentrations of multiple ligands by a collection of promiscuous receptors. We demonstrate that crosstalk can improve precision in concentration sensing and discrimination tasks. 
To achieve superior precision, the additional information about ligand concentrations contained in short binding events of the non-cognate ligand should be exploited. We present a proofreading scheme to realize an approximate estimation of multiple ligand concentrations that reaches a precision close to the derived optimal bounds. Our results help rationalize the observed ubiquity of receptor crosstalk in molecular sensing.
\end{abstract}

\maketitle

\section{Introduction}

Living cells need to collect information with high precision to respond and adapt to their environment \cite{Bowsher2014}. For example, chemotactic swimming bacteria can react to changes in concentrations of nutrients and toxins \cite{Wadhams2004}; and cells from the innate immune system can recognize distinct microbial components and initiate immune responses \cite{Akira2006}. Presence or concentrations of key ligands are measured via receptor proteins, which are usually located in the cell surface, and later processed by complex downstream signaling networks to trigger cellular responses. The accuracy of these measurements suffer from multiple sources of noise, including the transport of ligands by diffusion, the binding of the ligands to the receptors after they have arrived to the surface, and the communication between components of the signaling network.

In recent years, the fundamental limits to cellular sensing has received thorough theoretical consideration \cite{tenWolde2016,Bialek2005,Kaizu2014,Aquino2016}. Berg and Purcell \cite{Berg1977} were the first to study the problem and found that the sensing error can be minimized by either increasing the number of receptors or the number of measurements per receptor. More recently, Endres and Wingreen showed \cite{Endres2009} using a maximum likelihood estimation that the accuracy can be increased by a factor of 2 by solely taking into account the unoccupied time intervals. Further theoretical work has concentrated on understanding the limits in cellular sensing for single receptors with spatial \cite{Rappel2008,Rappel2008b,Hu2010,Endres2008,Smith2015} and temporal gradients \cite{Mora2010,Lalanne2015}, for multiple receptors \cite{Skoge2011,Roob2016,Vijay2016} and even for cells that can communicate \cite{Fancher2017,Mugler2016,Varennes2017}. The thermodynamic cost \cite{Barato2014,Lang2014,Govern2014b,Lan2012,Mehta2012} and the trade-offs between different resources for sensing \cite{Govern2014,Becker2015,Siggia2013} has also been explored at large. 

Most of the aforementioned models assume that receptors sense individual ligands. However, cognate ligands usually reside among other spurious ligands and receptor must identify them accurately \cite{Lalanne2013,Lalanne2015,Francois2013,Francois2016}. Recently, Mora \cite{Mora2015} derived the fundamental limit to measuring concentrations among spurious ligands, and devised a signaling network, based on a kinetic proofreading scheme \cite{Hopfield1974}, to approximately reach this limit. Similarly, Singh and Nemenman \cite{Singh2017} found that a single receptor is capable of correctly measuring two different ligands with disparate binding affinities using a similar network. These studies focused on understanding the optimal sensing capacity of one receptor with a cognate and one (or many) non-cognate ligands. In reality, receptors with different cognate ligands may communicate with each other through their downstream signaling networks, thereby increasing the efficiency of the measurement of the concentration of all ligands. This \emph{crosstalk} between receptors is likely to be the case for systems with a larger number of ligands than receptors. For example, in the bone morphogenetic protein signaling pathway, more than 30 different ligands interact with only 7 different receptors \cite{Mueller2012,Antebi2017}. Similar circumstances arise for Toll-like receptors in the innate immune system \cite{Kawai2011} and T-cell receptors in the adaptive immune system \cite{Sewell2012,Mason1998}. The fundamental limits to how different receptors combine information through crosstalk is currently unknown. 

In this paper, we place physical limits on concentration sensing with crosstalking receptors. We consider two types of receptors, each with their own cognate ligands and compare specific to crossreactive binding. To gain intuition, we first consider specific receptors which never bind to non-cognate ligands. We then analyze the more realistic case where both receptors bind to both ligands with different binding strengths in a background with other ligands. In the latter case, we call the receptors specific if the off-rates of the non-cognate ligand are as high as those for the background ligands.
In both cases, we demonstrate that crosstalk outperforms specific binding of ligands in some parameter regimes as measured by relative errors on concentration estimation of the two ligands. We also discuss a related problem: detection of the presence of a ligand over a given concentration threshold using sequential probability ratio tests.

\section{Maximum likelihood analysis of two crosstalking receptors}

We consider a situation of crosstalk in the simplified case of two receptors, labeled $A$ and $B$, and two ligands, labeled $1$ and $2$. The binding rates of the two receptors $k_{A}$ and $k_{B}$ are assumed to be independent of the identity of the ligand, as in the case of diffusion-limited binding, in which case $k_{A} = 4Ds_{A}, k_{B} = 4Ds_{B}$, where $s_{A},s_{B}$ are the sizes of idealized circular receptors located on the cell surface, and $D$ is the diffusivity of the ligand molecules. The distinction between the two ligands appears in the distinct rates of unbinding, which are denoted by $r_{{A},1},r_{{A},2}, r_{{B},1}, r_{{B},2}$. In the presence of both ligands, receptors alternate between bound and unbound states with exponential waiting times associated with the off and on rates respectively. In general, the mean occupancy of each receptor contains useful information about the concentrations of each ligand. However, the temporal information contained in the sequence of bound and unbound times is lost; a maximum likelihood estimation based off the binding and unbinding events yields an estimator that is unbiased and asymptotically achieves the least variance, as given by the Cram\'{e}r-Rao bound \cite{Kay2001,Endres2009,Mora2015}. For independent receptors, we may split the log-likelihood $\mathcal{L}$ into contributions from binding events at each receptor, $\mathcal{L} = \mathcal{L}_{A} + \mathcal{L}_{B}$. For ligands with concentrations $c_1$ and $c_2$ (with $c_{\text{tot}} = c_1 + c_2$), the probability of a sequence of bound and unbound times on receptor $A$ is written as

\begin{figure}
\includegraphics[width=\linewidth]{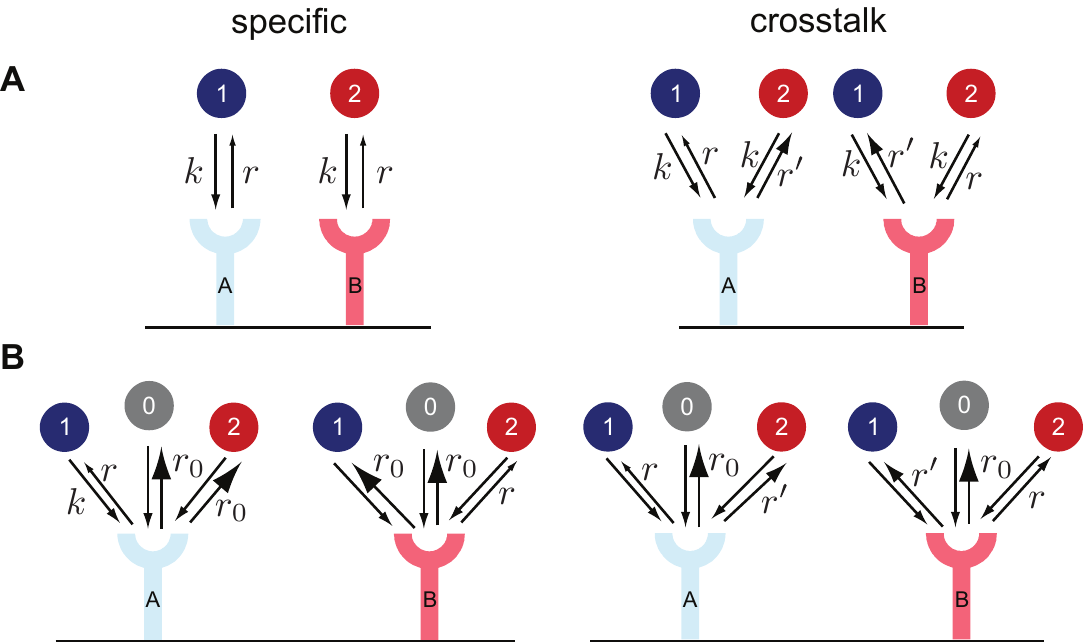}
\caption{{\bf Specific and crosstalk receptors for concentration sensing}. (A) The receptors A and B bind to their cognate ligand 1 and 2 respectively with the same rate $k$. Crosstalk is defined as the receptors additionally binding to another non-cognate ligand. The non-cognate ligand unbinds faster ($r'>r$), which allows for discrimination between the two types of binding events. (B) The two receptors bind to both ligand species, as well as to a pool of background ligands 0, all with the same binding rate $k$. The off rates of the non-cognate ligand can either be identical to the background ligands (specific case, $r<r'=r_0$) or in between the background and the cognate ligand (crosstalk case, $r<r'<r_0$).
\label{fig:schematic}}
\end{figure}

\begin{widetext}
\begin{align}
P(\{u_i^{A}, b_i^{A}\}) \equiv e^{\mathcal{L}_{A}} = \prod_{i=1}^{n_{A}} e^{-k_{A} c_{\text{tot}} u_i^{A}}\left[k_{A} c_1 r_{{A},1} e^{-r_{{A},1}b_i^{A}} + k_{A} c_2 r_{{A},2}e^{-r_{{A},2}b_i^{A}} \right] ,
\end{align}
where $\{u_i^{A}, b_i^{A}\}$ denotes the sequence of unbound and bound times, and the index $i$ runs from one to the total number of binding events $n_{A}$ in a fixed time interval $T$, which is taken to be much longer than the typical binding and unbinding times. A similar expression can be written for receptor $B$. The log-likelihood now reads
\begin{align}\label{eq:loglikelihood}
\mathcal{L} = \sum_{R = A,B} \left\{ -k_{R}c_{\text{tot}}T_{u}^{R} + n_{R}\log k_{R} c_{\text{tot}} + \sum_{i=1}^{n_{R}} \log\left( x r_{{R},1}e^{-r_{{R},1}b_i^{R}} + \left(1-x\right)r_{{R},2}e^{-r_{{R},2}b_i^{R}} \right) \right\},
\end{align}
with $x = c_1/c_{\text{tot}}$ and $T_u^{R}$ is the total unbound time of the receptor $R$. Maximum likelihood (ML) estimates of $x$ and $c_{\text{tot}}$ are obtained from the conditions $\frac{\partial \mathcal{L}}{\partial x}\rvert_{x^*} = 0$ and $\frac{\partial \mathcal{L}}{\partial c_{\text{tot}}}\rvert_{c_{\text{tot}}^*} = 0$. The ML estimate of the total concentration, $c_{\text{tot}}^*$, is given by
\begin{align}\label{eq:mlctot}
c_{\text{tot}}^* = \frac{n_{A} + n_{B}}{k_{A}T_u^{A} + k_{B}T_u^{B}}
\end{align}
The ML estimate $x^*$ satisfies the equation
\begin{align}\label{eq:mlx}
\sum_{i=1}^{n_{A}} \frac{\alpha_Ae^{-\left(\alpha_A - 1\right)r_{A,2}b_i^{A}} - 1}{x^*\alpha_Ae^{-\left(\alpha_A - 1\right)r_{A,2}b_i^{A}} + \left(1 - x^* \right)} - \sum_{i=1}^{n_{B}} \frac{\alpha_Be^{-\left(\alpha_B - 1\right)r_{B,1}b_i^{B}} - 1}{\left(1 - x^* \right)\alpha_Be^{-\left(\alpha_B - 1\right)r_{B,1}b_i^{B}} + x^*} = 0.
\end{align}
Here, we have defined $\alpha_A = r_{A,1}/r_{A,2}$ and $\alpha_B = r_{B,2}/r_{B,1}$. 
\end{widetext}

\section{Precision in concentration sensing}

Since $c_{\rm tot}$ and $x$ are involved in separate terms of the log-likelihood (\ref{eq:loglikelihood}), the variances of their optimal estimators are given, in the limit of large numbers of binding events, by the inverse of their respective Fisher information, $\delta x^2=\langle \frac{\partial^2 \mathcal{L}}{\partial x^2} \rangle^{-1}$ and $\delta c_{\rm tot}^2=\langle \frac{\partial^2 \mathcal{L}}{\partial c_{\rm tot}^2} \rangle^{-1}$,
where the angled brackets denote an expectation over the distribution parametrized by the true parameter values.
A similar approach is taken in \cite{Mora2015}; here, we simply write the final form for the variance in the estimates of $c_{\text{tot}}$ and $x$:
\begin{align}\label{eq:errors}
\delta c_{\text{tot}}^2 &= \frac{c_{\text{tot}}^{2}}{\langle n_A \rangle + \langle n_B \rangle}, \\ \label{eq:errordx}
\delta x^2 &= \left(\frac{\langle n_A \rangle}{f\left( x, \alpha_A\right)} + \frac{\langle n_B \rangle}{f\left( 1- x, \alpha_B\right)}\right)^{-1}\text{, where} \\
f\left( x, \alpha\right)^{-1} &= \int_0^{\infty} dt e^{-t}\frac{\left( \alpha e^{-\left(\alpha - 1\right)t} - 1\right)^2}{x \alpha e^{-\left(\alpha - 1\right)t} + \left(1 - x \right)}, \label{eq:f}
\end{align}
where averages and variances are taken over realizations of sequences of bound and unbound times (or equivalently, averaged over one sequence in the limit of large $T$) and $\langle n_i \rangle= T/((k_i c_{\text{tot}})^{-1} +x r_{i,1}^{-1}+(1-x)r_{i,2}^{-1})$.
Before proceeding further, we make a few simplifying assumptions of symmetric binding of the two ligands on the two receptors. Particularly, we assume $r_{A,1} = r_{B,2} = r$, $r_{A,2} = r_{B,1} = r'$ and $k_A = k_B = k$, which give $\alpha_A = \alpha_B \equiv \alpha=r/r'$. For $\alpha < 1$, the conditions imply that ligand 1 acts as a cognate ligand for receptor A and a non-cognate ligand for receptor B, whereas the reciprocal relationship is true for ligand 2. The discriminability of the two ligands is set by $\alpha$, which measures the ratio of bound times of the cognate and non-cognate ligands. It is convenient to nondimensionalize the concentrations of the two ligands as $\tilde{c}_1 = kc_1/r$ and $\tilde{c}_2 = kc_2/r$. The mean number of binding events on the two receptors is given by the total time divided by the mean time for each binding and unbinding cycle, which gives
\begin{align}
    \langle n_A \rangle = \frac{rT\left(\tilde{c}_1 + \tilde{c}_2 \right)}{1 +\tilde{c}_1 + \alpha\tilde{c}_2 }, \quad \langle n_B \rangle = \frac{rT\left(\tilde{c}_1 + \tilde{c}_2 \right)}{1 + \alpha \tilde{c}_1 + \tilde{c}_2 }.
\end{align}

In the case of specific receptors i.e., when each ligand binds to only one receptor type, the minimum variance of the estimated concentration of each ligand can be derived in a similar fashion from the log-likelihood  (see \cite{Endres2009}). If we suppose ligand 1 binds specifically to receptor A (with the same binding and unbinding rates as in the crosstalk case above), and ligand 2 binds only to receptor B, the error in ML estimation is given by
\begin{align}
    \delta c_1^2 = c_1^{2}/\langle n_A \rangle_{\text{spec}}, &\quad \delta c_2^{2} = c_{2}^2/\langle n_B \rangle_{\rm spec}\text{, where} \\
    \langle n_A \rangle_{\text{spec}} = \frac{rT\tilde{c}_1}{1+\tilde{c}_1}, &\quad \langle n_B \rangle_{\rm spec} = \frac{rT\tilde{c}_2}{1+\tilde{c}_2}, \label{eq:error_specific}
\end{align}
where, $\langle n_A \rangle_{\text{spec}}, \langle n_B \rangle_{\rm spec} $ are the average number of binding events for the specific receptor case in the same interval $T$. Note that \eqref{eq:error_specific} does not correspond to the $r' \to \infty$ limit in \eqref{eq:errors} and \eqref{eq:errordx}: when $r'$ is very large, non-cognate ligand bound times can be read easily because there is no cutoff for the readout of small bound times. This biological inconsistency can be removed by taking into account binding of non-specific molecules (see Section \ref{sec:nonspecific}).

To make a comparison between the effectiveness of crosstalking and specific receptors, it is more pertinent to estimate relative errors, $\delta c_1/c_1,\delta c_2/c_2$, as concentrations can span many orders of magnitude. In the limit of long times, where errors are Gaussian-distributed, the covariance matrix
\begin{equation}\label{eq:Sigma}
\Sigma=\left[\begin{array}{cc}\frac{\delta c_1^2}{c_1^2} & \frac{\delta c_1\delta c_2}{c_1c_2}\\ \frac{\delta c_1\delta c_2}{c_1c_2} & \frac{\delta c_2^2}{c_2^2}\end{array}\right]
\end{equation}
describes the magnitude and shape of relative estimation errors as an ellipse around the true value. Its determinant, $\Omega=\mathrm{det}(\Sigma)$, gives the volume of that ellipse and is used as a measure of error. Intuitively, the discriminability between two pairs of concentrations $(c_1,c_2)$ and $(c_1',c_2')$ depends on the overlap in the areas of the two ellipses centered around the pairs \cite{Averbeck2006}.
In the crosstalk case, from \eqref{eq:errors} we have
\begin{equation}
\begin{split}
    \Omega_{\text{CT}} =& \frac{1}{x^2\left(1-x \right)^2}\frac{1}{\langle n_A \rangle + \langle n_B \rangle}\\
    &\times \frac{1}{\langle n_A  \rangle f\left(x,\alpha\right)^{-1}  + \langle n_B \rangle f \left(1-x,\alpha\right)^{-1}},
    \end{split}
\end{equation}
For the case of specific binding, the measurements of $c_1$ and $c_2$ are independent, and from \eqref{eq:error_specific}, the determinant of the covariance matrix is 
\begin{align}
    \Omega_{\text{S}} = \frac{1}{\langle n_A \rangle_{\text{spec}} \langle n_B \rangle_{\text{spec}}}. 
\end{align}

\begin{figure*}[t!]
\includegraphics[width=.8\linewidth]{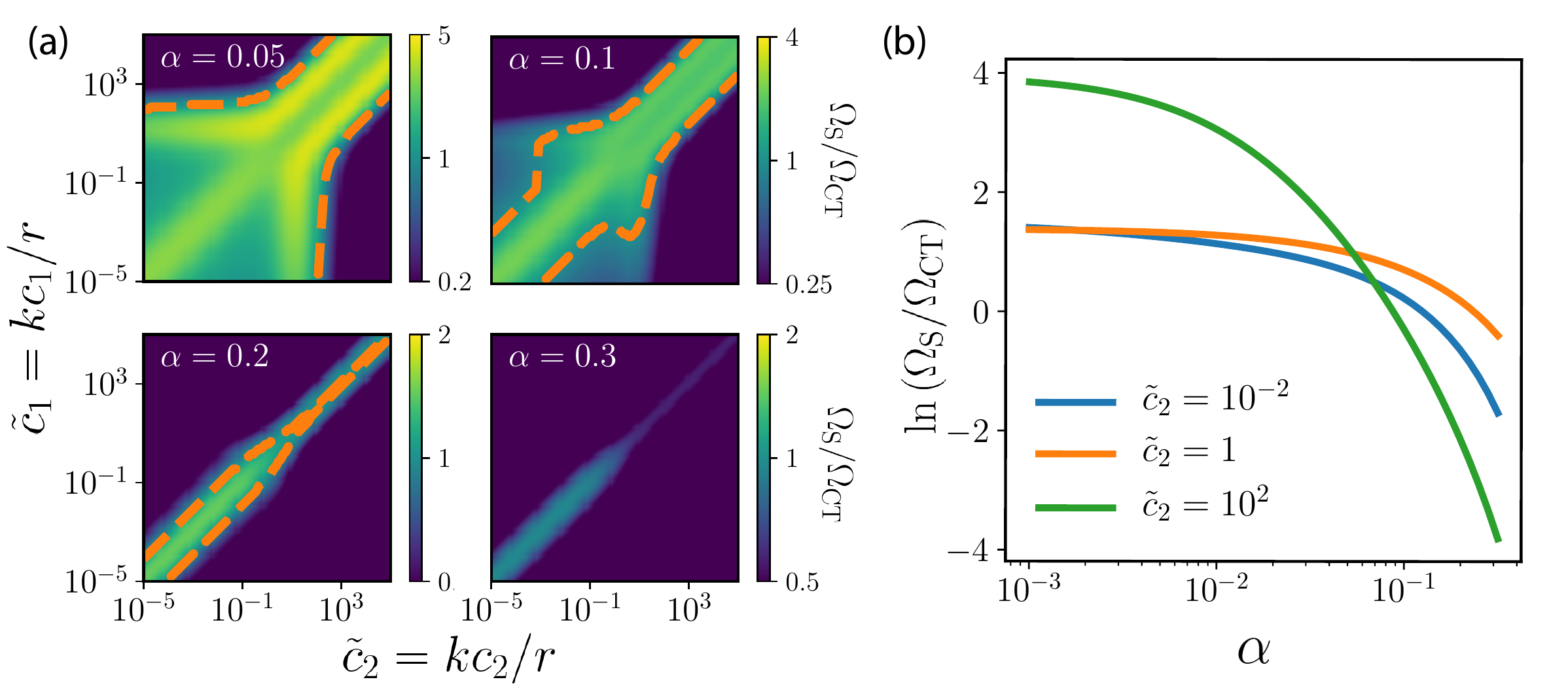}
\caption{{\bf Comparison of concentration sensing accuracy with specific and crosstalking receptors}. (a) Ratio between the specific and cross-talk errors, $\Omega_{\text{S}}/\Omega_{\text{CT}}$, for four different values of $\alpha$ and different concentration pairs. The region where crosstalk shows greater precision $\Omega_{\rm CT}<\Omega_{\rm S}$ is delineated by the orange dashed line. For $\alpha > \alpha_c \approx 0.27$, crosstalk does not exhibit greater precision in any range of concentrations. (b) The logarithm of the same ratio for $\tilde{c}_1 = 1$ and three values of $\tilde{c}_2$ as a function of $\alpha$.} 
\label{fig:phaseDiagrams}
\end{figure*}

In the limit of weak crosstalk, $\alpha \ll 1$, we find that crosstalk between receptors always improves sensing capacity over non-crosstalking receptors, regardless of the concentrations of each ligand. To see this, note $f(x,\alpha) \approx (1-x)$ and $f(1-x,\alpha) \approx x$ (from \eqref{eq:f}), and so
\begin{align}
\label{eq:omega_ratio}
\Omega_{\text{S}}/\Omega_{\text{CT}}\approx 2 \frac{(\mu +2 x(1-x))(\mu + 1/2)}{(\mu+x)(\mu+1-x)} > 1,
\end{align}
where $\mu = (\tilde{c}_1 + \tilde{c}_2)^{-1}$.
In particular, when $c_1 \approx c_2$, the number of binding events for the crosstalking receptors is twice that for the specific receptors and consequently, $\Omega_{\text{S}}$ is a factor two greater than $\Omega_{\text{CT}}$ in this limit. In Figure \ref{fig:phaseDiagrams} we mark the regions in the $(\tilde c_1,\tilde c_2)$ plane where $\Omega_{\text{S}} > \Omega_{\text{CT}}$ for different values of $\alpha$. We make two major observations. First, for $\alpha = 0$ i.e., for perfect discriminability, crosstalking receptors always show lower estimation error (Eq. \ref{eq:omega_ratio}). Second, there is a critical value of $\alpha$, $\alpha_c \approx 0.27$, beyond which $\Omega_{\text{S}}$ is smaller than $\Omega_{\text{CT}}$ for all concentrations. Thus, the usefulness of crosstalk diminishes with increasing $\alpha$, as in Ref.~\cite{Singh2017}.

\section{Performance comparison in discrimination tasks}

\begin{figure}[t!]
\begin{center}
\includegraphics[width=.8\linewidth]{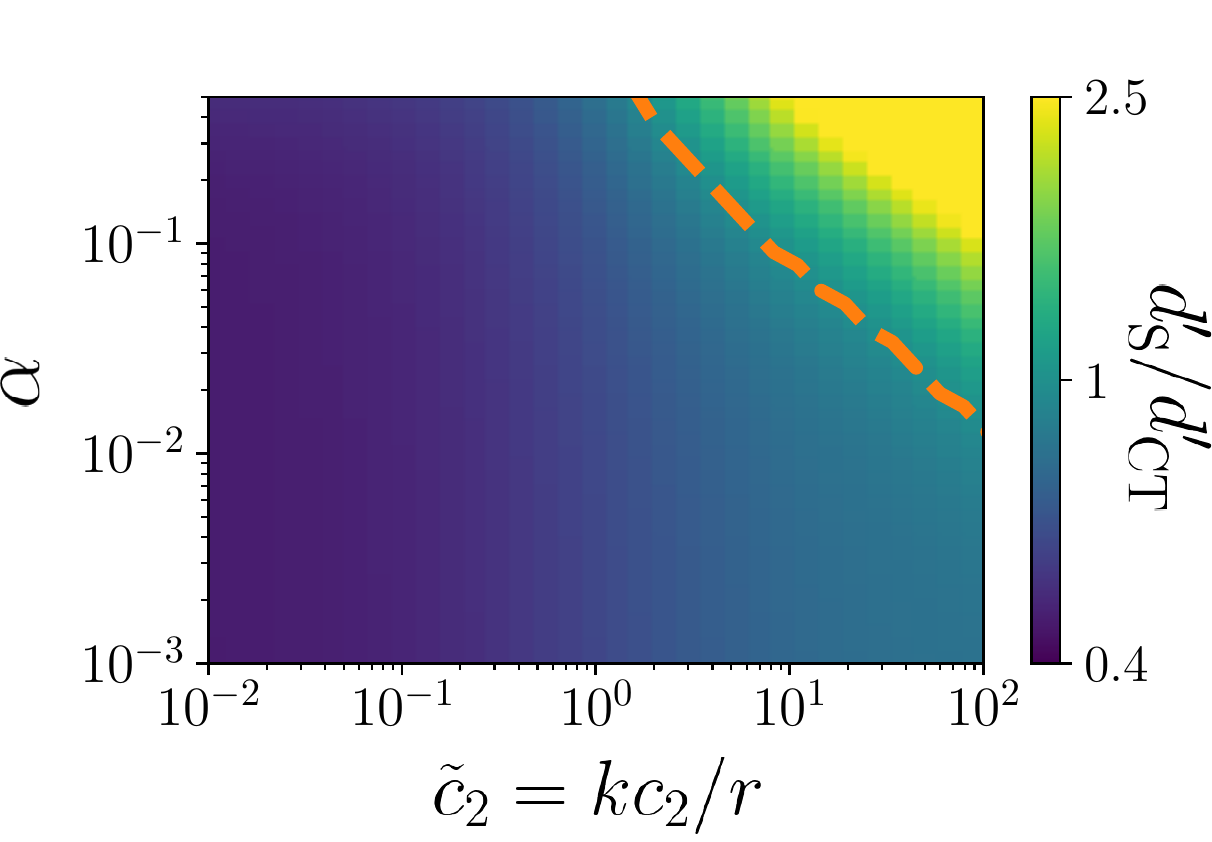}
\caption{{\bf Crosstalking receptors show improved discriminability relative to specific receptors.} Ratio of sensitivity index $d'$ for the cross-talk ($d'_{\rm CT}$) and specific ($d'_{\rm S}$) cases, as a function of the concentration of the second ligand $c_2$, and the specificity ratio $\alpha=r/r'$, in the task of discriminating two values of the first ligand concentration, $kc_1/r=0.5$ or $kc'_1/r=2$.
The orange line separates the regions in which each of the two strategy (crosstalk or specific) is optimal. Crosstalk is beneficial for small $c_2$ and $\alpha$, i.e. when the ligands are easy to distinguish and do not saturate the receptor.
The sensitivity index was computed using $rT = 10^4$.}
\label{fig:discriminability}
\end{center}
\end{figure}
Next, we consider the task of discriminating between two external environmental states or 'hypotheses', $H_0$ and $H_1$, corresponding to ligand concentrations $(c_1,c_2)$ and $(c_1',c_2')$ respectively. A Bayes-optimal decision-maker compares the likelihood ratio, or equivalently, the difference in log-likelihoods, $\mathcal{L}_0 - \mathcal{L}_1 \equiv \Delta \mathcal{L}$, for the two hypotheses given the observed bound and unbound times and makes a decision if $\Delta \mathcal{L}$ crosses a certain threshold $\theta$ in a given amount of time $T$. The false positive and false negative error rates depend on $\theta$ and the distribution of $\Delta \mathcal{L}$ under the hypotheses. To conveniently compare the discriminability for the crosstalk and specific cases, we use the discriminability (or `sensitivity index' $d'$), defined as 
\begin{align}
d' = \frac{\langle \Delta \mathcal{L} \rangle_c - \langle \Delta \mathcal{L} \rangle_{c'}}{\sqrt{\langle \delta^2 \Delta \mathcal{L} \rangle_c + \langle \delta^2 \Delta \mathcal{L} \rangle_{c'}}},
\end{align}
where subscripts $c$ and $c'$ denotes expectation values under concentrations $(c_1,c_2)$ and $(c'_1,c'_2)$ respectively.
If $\Delta \mathcal{L}$ is Gaussian-distributed and its variance equal for both sets of concentrations, $d'$ and $\theta$ together uniquely determine the false positive and false negative error rates. 
For large $T$, the central limit theorem guarantees that the Gaussian approximation is a good one. Although the variances $\langle \delta^2 \Delta \mathcal{L} \rangle_c$ and $\langle \delta^2 \Delta \mathcal{L} \rangle_{c'}$ are not equal in general, $d'$ is often used as a general measure of discriminability.

To calculate the mean and variance of $\langle \Delta \mathcal{L} \rangle_c$, we first observe that since the two receptors are independent, it is sufficient to compute them for a single receptor and sum them up. We show in the Appendix that for a single receptor with arbitrary unbound and bound time distributions, 
the cumulant generating function of $\langle \Delta \mathcal{L} \rangle_c$ has a simple form in the limit of large $T$, from which we derive explicit forms for the mean and variance. 
In Figure \ref{fig:disc_formula}, we validate these analytical expressions using numerical simulations. 

The discriminabilities for the crosstalking and specific receptors, $d'_{\text{CT}}$ and $d'_{\text{S}}$, are compared in Figure \ref{fig:discriminability}. The orange line mark the region where crosstalk offers greater discriminability compared to the specific case. Large concentrations of the non-cognate ligand can mask the accurate discrimination between different concentrations of the cognate ligand, as evidenced by the blue region. As $\alpha \to 0$, masking plays a limited role, since even though binding events are dominated by the non-cognate ligand, the bound times are extremely short and easily distinguishable from the bound times of the cognate ligand. 

 % nonspecific_pool.tex
\section{Concentration estimates of two ligands in a pool of nonspecific ligands}\label{sec:nonspecific}

\begin{figure*}
\centering{\includegraphics[width=.8\linewidth]{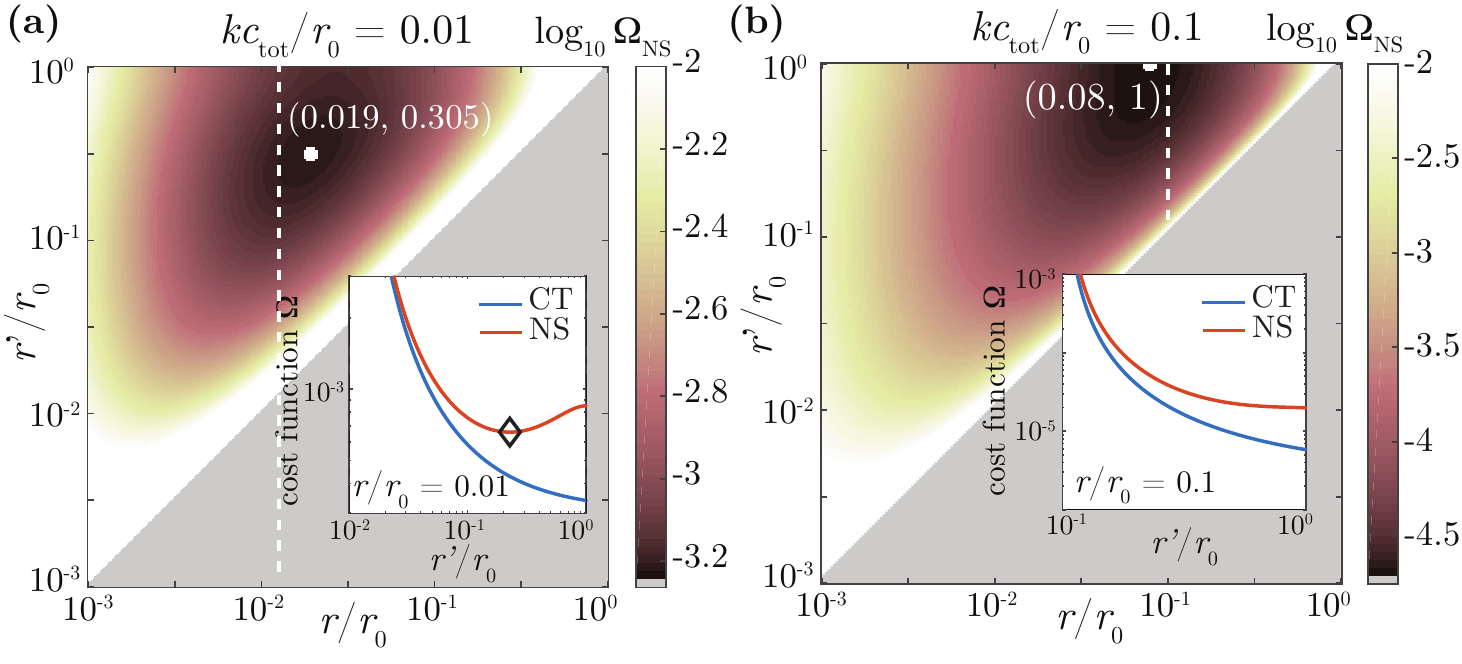}}
\caption{\textbf{Optimal unbinding rates in presence of background unspecific binding.}
Logarithm of the error, $\log\Omega_{NS}$ as a function of the
the unbinding rate of the cognate ligands, $r$, and the unbinding of noncognate
ligands, $r'$, for the ML estimates in the crosstalk model with nonspecific ligands with $x=y=0.4$. (a) Weak background binding ($kc_{\rm tot}/r_0=0.01$). $\Omega_{NS}$
attains a local minimum at $r_{\min}<r'_{\min}<r_0$,
meaning that crosstalk optimizes the accuracy of
concentration sensing. In the inset, we plot $\Omega_{NS}$ (red) and $\Omega_{CT}$
(blue) as a function of $r'/r_0$, at fixed $r/r_0=0.01$ (along the dotted
white line in the main plot). While
$\Omega_{CT}$ decreases monotonously as $r'$ increases (and $\alpha$
decreases), $\Omega_{NS}$ attains a local minimum (black diamond).
A good estimation of the cognate ligand concentrations requires all
unbinding rates to be as dissimilar as possible, ensuring that the identity of the bound
ligand can be faithfully inferred from the binding times. (b) Strong background binding ($kc_{\rm tot}/r_0=0.01$). The minimum is reached at the boundary,
$(r'_{\min}=r_{0})$, meaning that the noncognate ligand is indistinguishable from background binding and thus treated as noise. This optimal solution thus reduces to the case of specific binding in the presence of background binding. Plots are obtained with $kc_{1}/r_0=0.4$, $kc_{2}/r_0=0.4$, $kc_{3}/r_0=0.2$, and $r_0T=10^{4}$. \label{fig:Omega_vs_r_rp}}
\end{figure*}

Next, we consider a more realistic scenario the cell faces: the problem
of concentration estimation of two ligands in a presence of a pool
of background nonspecific ligands. This scenario also allows us to resolve the
inconsistency of the limit of perfect specificity, $\alpha\to 0$. As we observed earlier, that limit did not reduce to the case of no cross-talk, because of infinitely short but mathematically informative nonspecific binding events. Adding a background of nonspecific ligands removes the informative content of these spurious events by making them indistinguishable from background ligand binding.
Here, as before, we have two receptors,
$A$ and $B$, and two cognate ligands, $\text{1}$ and $2$, whose
concentrations $c_{1}$ and $c_{2}$ the cells needs to be estimate.
In addition, there exists a pool of nonspecific ligands labeled $0$,
with concentration $c_{0}$. The receptors now have to differentiate
between three types of ligands to correctly estimate their concentrations.
We assume diffusion-limited ligand-receptor binding as before. The off-rates for the nonspecific ligands are $r_{A,0}=r_{B,0}=r_0\geq r'>r$ from both receptors. With
total concentration given by $c_{\rm tot}=c_0+c_{1}+c_{2}$, and relative fractions $x=c_1/c_{\rm tot}$, $y=c_2/c_{\rm tot}$, the probability
$P(\{u_{i}^{R},b_{i}^{R}\})$ of a sequence of bound and unbound times can be written as a function of the on-rate,
various off-rates and concentrations. As before, the ML estimate of $\bm{{\theta}}=[x,y,c_{\rm tot}]$
can be obtained by maximimizing the log-likelihood $\mathcal{L}=\log P$, by setting $\frac{\partial\mathcal{L}(\bm{\theta})}{\partial\bm{\theta}}\rvert_{\bm{\theta}^{*}}=0$.
Further, in the limit of large numbers of binding events, the Cram\'{e}r-Rao
bound guarantees that the covariance matrix of the estimator $\bm{\theta}$
is given by the inverse of the Fisher Information matrix:
\begin{equation}
\langle \delta\bm{\theta}^T\delta \bm{\theta}\rangle=
-\left.\Big<\dfrac{\partial^{2}\mathcal{L}}{\partial\bm{\theta}^T\partial\bm{\theta}}\Big\rangle\right\rvert_{\bm{\theta}^{*}}^{-1}.
\end{equation}

The covariance matrix $\Sigma_{\text{NS}}$ of the relative errors on concentrations, $\delta c_1/c_1$ and $\delta c_2/c_2$, can then be obtained by a change of variable from $\bm{\theta}$ to $(c_0,c_1,c_2)$, from which the error volume $\Omega_{\text{NS}}=\det(\Sigma_{\text{NS}})$ is computed, analogous to $\Omega_S$ and $\Omega_{\text{CT}}$ defined earlier.

We are interested
in how $\Omega_{\text{NS}}$ varies as a function of $r$ and $r'$, to understand
if crosstalk (defined now as $r'<r_0$ by contrast to background unspecific binding, $r'=r_0$) can result in relatively better concentration estimates.
Up to scalar scaling, $\Omega_{\text{NS}}$ depends
only on the ratios $kc_{\rm tot}/r_{0}$, $r/r_{0}$ and $r'/r_{0}$ and
the ligand concentration fractions $x$ and $y$.

\begin{figure*}
\centering{\includegraphics[width=.8\linewidth]{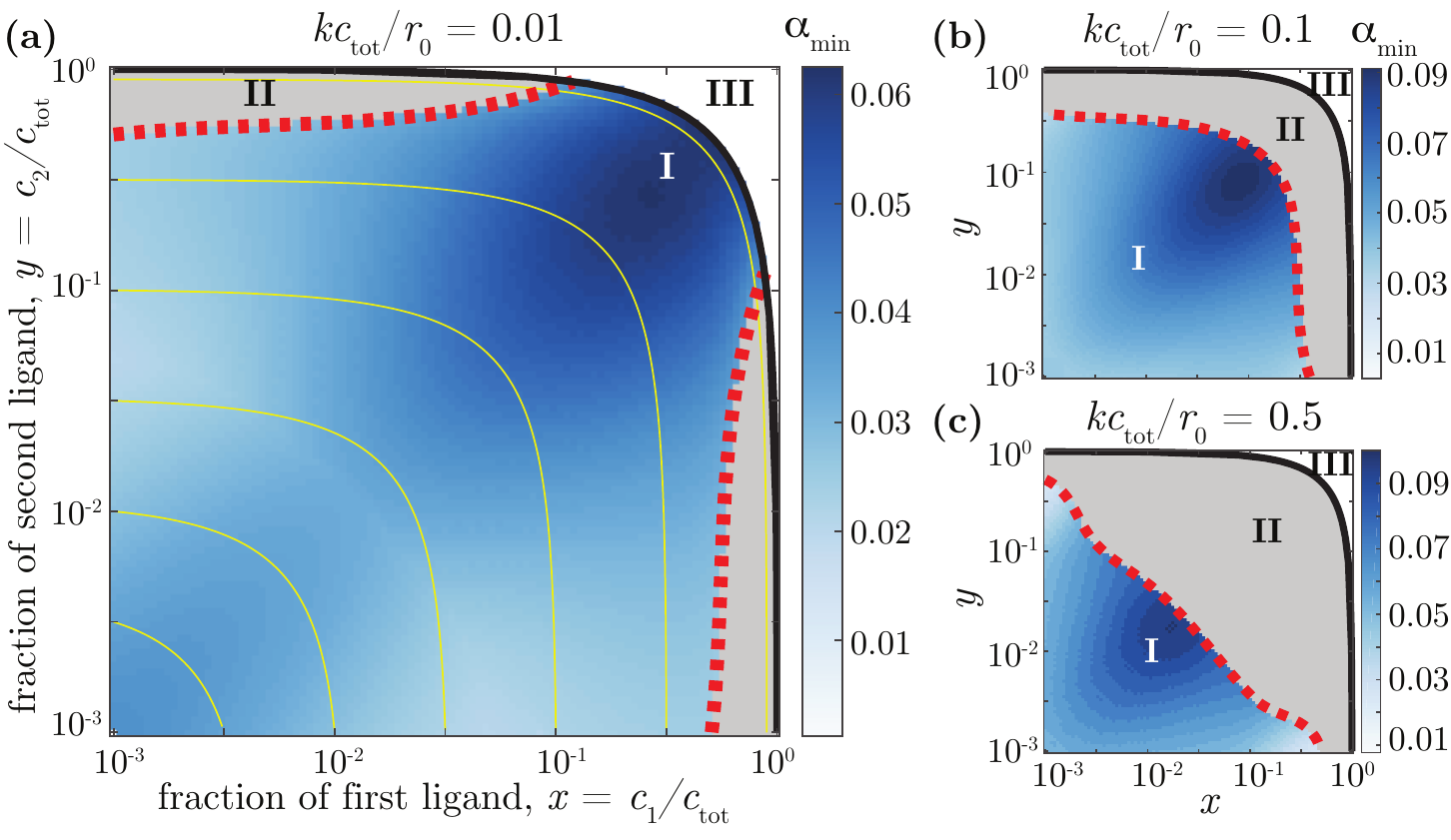}}
\caption{\textbf{Optimal strategy as a function of ligand fractions.}
Three phases are shown:
I, crosstalking receptors are optimal ($r_{\min}<r'_{\min}<r_{0}$);
II, specific receptors are optimal ($r_{\rm min}<r'_{\min}=r_{0})$;
III, impossible region ($x+y>1$).
The heatmap shows the optimal value of the specificity ratio, $\alpha_{\rm min}=r_{\rm min}/r'_{\rm min}$. Yellow lines indicate a constant $c_0$ (or $x+y$).
The phase diagram depends on the strength of background binding: (a) $k c_{\rm tot}/r_0 = 0.01$, (b) $k c_{\rm tot}/r_0 = 0.1$, and (c) $k c_{\rm tot}/r_0=0.5$. As this strength increases, the region where specific binding is optimal (phase II) extends towards smaller values of $x$ and $y$.
\label{fig:alphamin}}
\end{figure*}

The estimation error quantified by $\Omega_{\text{NS}}$ attains a minimum
(Fig. \ref{fig:Omega_vs_r_rp}a) at some finite values of $r=r_{\min}$
and $r'=r'_{\min}<r_{0}$, meaning that a finite crosstalk minimizes
the estimation error. At a fixed $r$, for $r'>r$, while $\Omega_{\text{CT}}$
monotonically decreases with increasing $r'$, $\Omega_{\text{NS}}$ is non-monotonic
with a local minimum at $r'=r'_{\min}\in(r,r_{0})$ (inset, Fig. \ref{fig:Omega_vs_r_rp}a).
The existence of a minimum results from a tradeoff between three conflicting effects.
First, when $r'$ is large, receptors cannot reliably distinguish between
the noncognate and the nonspecific ligand, resulting in an increased
estimation error for $x$ and $y$. On the other hand, larger values
of $r'$ and $r$ result in more binding events, thereby improving
statistics and accuracy. Lastly, $r$ and $r'$ should as far as possible from each other for each receptor to be able to distinguish cognate from noncognate ligands.
These opposing forces result in
a local minimum at $(r_{\min},r'_{\min})$, balancing the need for speed (high unbinding rates $r$, $r'$) with that of specificity (low but different $r$ and $r'$), ensuring that receptors can maximally
distinguish between the three different kinds of ligands by looking
at their binding times.

Distinct regimes emerge depending on the value of the dimensionless parameter $kc_{\rm tot}/r_{0}$, which can be viewed as an effective unspecific receptor occupancy quantifying the effect of background binding, as well as on the ligand fractions $x$ and $y$.
For small $kc_{\rm tot}/r_{0}$,  (Fig. \ref{fig:Omega_vs_r_rp}a), a tradeoff exists between speed and specificity ($r'_{\rm min}<r_0$), while 
for large $kc_{\rm tot}/r_{0}$ (Fig. \ref{fig:Omega_vs_r_rp}b), the low receptor availability caused by unspecific background binding makes the speed requirement dominate, resulting in absence of cross-talk in the optimal solution ($r'_{\rm min}=r_0$).
Fig. \ref{fig:alphamin} shows the phase diagram as a function of the ligand fractions $x$ and $y$, for fixed values of $kc_{\rm tot}/r_{0}=0.01$. Three phases emerge: cross-talk ($r_{\rm min}<r_0$, I), no cross-talk ($r_{\rm min}=r_0$, II), and impossibility region ($x+y>1$, III).

When one ligand is present in very low concentrations and the other in high concentrations (low $x$, high $y$, or vice-versa), introducing cross-talk would cause the abundant ligand to saturate both receptors, keeping the receptor that is cognate to the sparse ligand from sensing its concentration. In that regime, cross-talk is not optimal (region II of Fig.~\ref{fig:alphamin}a).
As $kc_{\rm tot}/r_{0}$ increases (Fig. \ref{fig:alphamin}b),
this region spreads towards smaller values of $x$ and $y$. Yet even
when the effect of background ligand is felt strongly, cross-talk is still advantageous when the cognate ligands are sparse ($x+y\ll 1$).

\section{A biochemical network scheme that reaches close to the optimal bounds}

\begin{figure*}
\begin{center}
\includegraphics[width=0.8\linewidth]{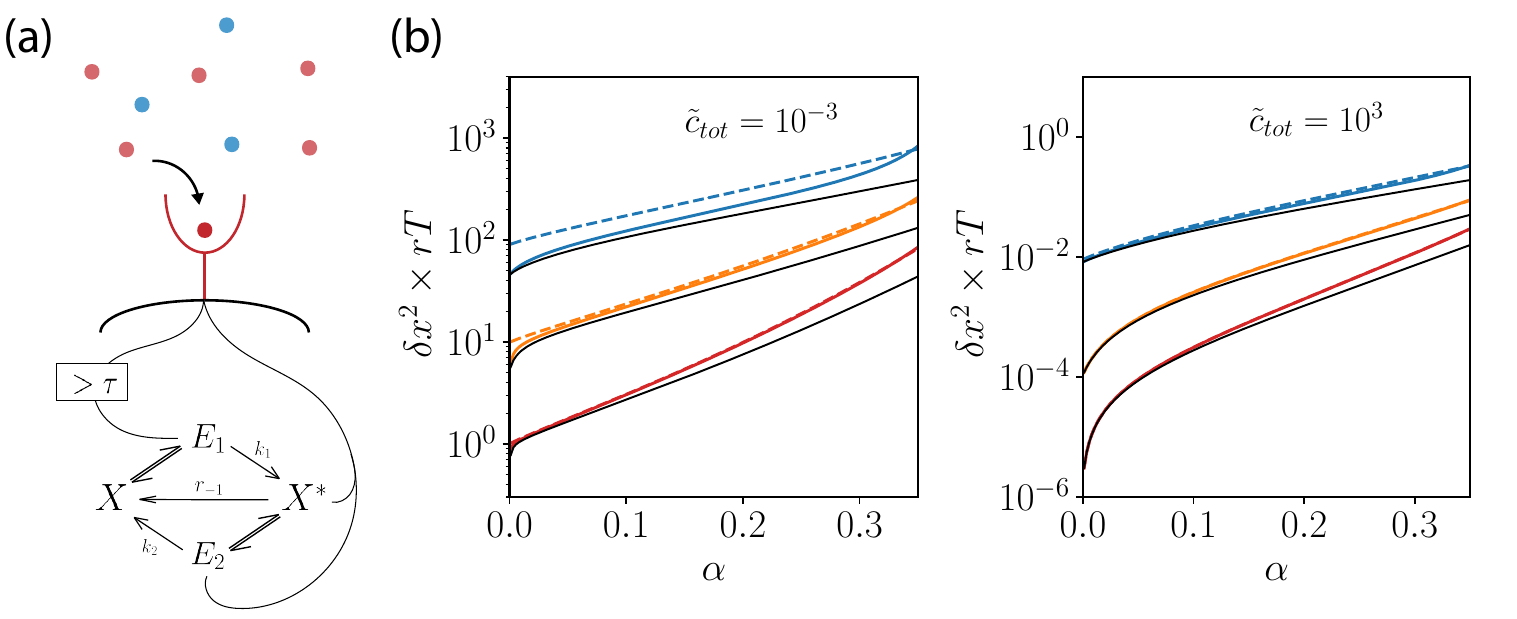}
\caption{(a) Network scheme to estimate $x=c_1/c_{\rm tot}$ using the fraction of binding events that last longer than some threshold $\tau$. An estimate of $x$ is read off from the fraction of $X$ molecules in the active state (see text). (b) The error in estimating $x$ for $x = 10^{-1},10^{-3}, 10^{-5}$ (blue, orange, red respectively) using \eqref{eq:xhat} (solid, colored lines) and the crude estimate of using only one receptor (dashed lines) is compared with the optimal ML error (solid, black lines). }
\label{fig:networkerror}
\end{center}
\end{figure*}

In this section, we present a simple kinetic proofreading-based scheme that implements an approximate maximum likelihood estimation with a precision close to the derived optimal bounds. In the following analysis, for simplicity, we revert to the former case of one cognate and one non-cognate ligand. We note that the ML estimate for the total concentration $\hat{c}_{\text{tot}}  = ({n_{A} + n_{B}})/({k_{A}T_u^{A} + k_{B}T_u^{B}})$ has a simple expression; the terms in the numerator and denominator can be measured biochemically and combined to form $\hat{c}_{\text{tot}}$. To estimate $x$, the scheme relies on a proofreading `classifier' associated to each receptor that distinguishes between bound times above and below a certain threshold. An unbiased estimate of $x$, which we denote by $\tilde{x}_R$, can be formed for each receptor $R$ ($R = A$ or $B$), based on the fraction $\hat{f}_R$ of binding events where the ligand is bound longer than $\tau_R$. Defining $y_R = e^{-r' \tau_R}$, $h_R = e^{-r \tau_R}-e^{-r' \tau_R}=y_R^{\alpha} - y_R$, this estimate reads:
\begin{align}
    \hat{x}_A = \frac{\hat{f}_A - y_A}{h_A} \\
    \hat{x}_B = \frac{y_B^{\alpha} - \hat{f}_B}{h_B}
\end{align}
The error in each estimate is $\delta x_R^2 = {f_R\left(1 - f_R\right)}/({h_R^2 n_R})$, where $n_R$ is the mean total number of binding events in time $T$ and $f_R \equiv \langle \hat{f}_R \rangle$. Because the estimates from receptors A and B stem from independent binding events,
the best way to combine them is through a weighted average of the estimates from each receptor:
\begin{align}\label{eq:xhat}
    \hat{x} = \beta \hat{x}_A + \left(1 - \beta \right) \hat{x}_B,
\end{align}
with $\beta = 1/\left(1 + \delta x_A^2/\delta x_B^2 \right)$. 

The value of the threshold $\tau_A$ can be optimized to yield the most precision in $\tilde{x}_A$. In Appendix B, we show that the optimum is reached for $r'\tau_A^* = (1-\alpha)^{-1} \log{\left(\frac{1-2\alpha}{\alpha x} \right)}$, valid for $\alpha < 1/2$. We note here that even though $\tau_A^*$ depends on $x$, the dependence is logarithmic. To implement adaptive thresholds that depend on $\log x$, we can imagine a collection of `gating' networks which apply different thresholds to the length of the binding events. An independent crude estimate of $x$ is sufficient to choose the network that has a threshold closest to $\tau_A^*$. The optimal value for $\tau_B$ has a similar form with $x$ replaced by $1-x$.

The cell can easily compute $\hat{x}_A$ and $\hat{x}_B$ using a proofreading motif followed by a downstream push-pull network as shown in Figure \ref{fig:networkerror}A. For concreteness, consider the case of receptor $A$. Suppose the proofreading motif produces a molecule of enzyme $E_1$ for each binding event longer than the threshold $\tau_A$. Further, every binding event generates a single protein $X$ and an enzyme molecule $E_2$. The enzymes $E_1$ and $E_2$ catalyze the conversion of $X$ to its active state $X^*$ and vice-versa, respectively. Assuming that $X$ and $X^*$ are in excess in the enzymatic reaction, the rate at which $X$ and $X^*$ are inter-converted is directly proportional to the enzyme numbers with catalytic rates denoted by $k_1$ and $k_2$. Suppose also that $X^*$ reverts to $X$ at a finite rate $r_{-1}$. At steady state, $dX^*/dt = 0$, and we have $r_{-1}X^* = k_1 E_1 - k_2 E_2$. Then, setting $k_1/r_{-1} = 1/h_A$ and $k_2/r_{-1} = y_A/h_A$, the fraction of $X$ molecules in the active state tracks $\hat{x}_A$.

Combining $\hat{x}_A$ and $\hat{x}_B$ as in \eqref{eq:xhat} using a biochemical network can be done at fixed $\beta$. For instance, when $x \sim 1/2$, the errors from both receptors are about the same and we may weight the estimate from each receptor equally, $\beta=1/2$.
However, tuning $\beta$ to reflect its dependency of the concentrations requires additional adaptive mechanisms.
In the regime $x\ll 1$, receptor A has the highest precision in estimating $x$, and $\tilde{x}_A$ can be taken as a crude estimate of $x$  (and symmetrically for B when $1-x\ll 1$).
In Figure \ref{fig:networkerror}B, we compare the error from this crude estimate against the optimal crosstalk error and the error from using the optimal network weights \eqref{eq:xhat}. This comparison shows that although the approximate biochemical solutions are not optimal, they stand reasonably close to the ML estimate.

\section{Discussion}

The key result of our paper is that crosstalk is generically the optimal strategy for sensing multiple ligand concentration using multiple receptors.
The theory predicts an optimal level of crosstalk, which balances the opposing requirements of maximizing the number of binding events with the receptor's ability to distinguish the ligands involved in those events.
For this discrimination to be possible, the difference between cognate and non-cognate binding affinities is maintained.

In the simplest theory that we presented (Sec. 2 and 3), higher unbinding rates are always advantageous because they increase receptor availability without hurting discriminability. However, in reality there are several reasons why high unbinding rates are not optimal. First, the molecular machinery required to process the information of the receptor binding state operates with its own incompressible time scale, which sets a lower bound on the duration of binding events that can be detected. Second, it is not realistic to assume that unbinding rate can be increased arbitrarily without affect binding rates as well. Very unspecific ligands that unbind very quickly are also less likely to bind in the first place, and the assumption of diffusion-limited binding rate is no longer valid. Lastly, for high enough unbinding rates, cognate and cross-talk binding events become indistinguishable from completely unspecific binding with other generic molecules. Modeling that situation as we did in Section \ref{sec:nonspecific} allowed us to find well-defined optimal unbinding rates.

In addition to maximizing the use of each receptor to gain information about each ligand, crosstalk has the additional advantage of expanding the dynamic range of concentrations over which ligands can be sensed. For instance, when a ligand is present at high concentration, its
cognate receptors will be fully saturated, making it difficult to reliably read off the concentration from the receptor's activity. Lower-affinity binding to a second receptor can then allow for more accurate sensing, as long as that receptor is not itself saturated by other ligands. 
In our language, when receptor $A$ is saturated by ligand 1, $c_1\gg r/k\equiv K_d$, then receptor $B$ is still sensitive to the concentration of ligand 1 in the regime $c_1\sim r'/k=\alpha^{-1}K_d$.
More generally in the presence of multiple ligands and multiple receptors, a good strategy could be to organize specificities (i.~e., unbinding rates) so that, for each ligand, dissociation constants collectively tile the sensory space. For such strategies to work, the concentration space must be sparse, meaning that only one or a few of the ligands of interest are present in large concentrations at the same time.

Cross-talk, also known as promiscuous binding, cross-reactivity, or multiplexing, depending on the context, is widespread in biology. It is an important feature of the BMP \cite{Mueller2012}, Notch, Wnt \cite{Wodarz1998}, and JAK-STAT \cite{Murray2007} pathways, as well as the Eph-ephrin system for cell positioning \cite{Jorgensen2009}, T- and B-cell receptor antigen recognition \cite{Mason1998,Mayer2015}, and olfactory receptors \cite{Hallem2006}. An often cited benefit of promiscuity is that it confers the ability to design combinatorial codes. In the context of olfaction \cite{Malnic1999}, such design can be advantageous in the presence of sparse odors \cite{Krishnamurthy2017}. Combinatorial codes also allow for flexible computations in signaling pathways \cite{Antebi2017}. In the adaptive immune system, cross-reactivity is necessary to cover the large space of possible antigens with a limited number of receptors \cite{Sewell2012,Mason1998}. Our results suggests another advantage of cross-talk: sensing accuracy. It would be interesting to study whether some of the biological systems that exhibit cross-talk make use of that benefit, and whether they are organized in a way that approaches the optimal solution.

{\bf Acknowledgements.} This work was initiated as a student project during the 2017 Carg\`ese school on biophysics held at the Institut d'\'Etudes Scientifique de Carg\`ese (Corsica, France), with support from the Centre National de la Recherche Scientifique through its training program and the {\em Groupement De Recherche International} ``Evolution, Regulation and Signaling'', PSL University, the National Science Foundation (NSF) Grant PHY-1740578, the Institute for Complex Adaptive Matter through NSF Grant DMR-1411344, and the Collectivit\'e Corse. IN was additionally partially supported by NSF Grant PHY-1410978.

\bibliographystyle{pnas}

\onecolumngrid

\appendix

\section{Cumulants of the log-likelihood for discrimination}
Consider a receptor switching between two states, bound and unbound, where the bound times $u$ and unbound times $b$ are drawn from general distributions that depend on the external ligand environment. We will consider the problem of discriminating two possible environmental states, labeled $H_0$ and $H_1$, given the time series of receptor states for a fixed time $T$, where $T$ is much larger than the typical bound and unbound times. Our interest is in calculating the cumulants of the log-likelihood difference between the two hypotheses, $\mathcal{L}_0 - \mathcal{L}_1 \equiv \Delta \mathcal{L}$, under the hypothesis that either $H_0$ or $H_1$ is the true external state. We shall use $P_i(u)$ and $Q_i(b)$ to denote the distributions of unbound and bound times respectively for $H_i$.  

The moment generating function $M(\lambda)$ of $\Delta \mathcal{L}$ under $H_i$ is
\begin{align}
M(\lambda) = \langle e^{\lambda \Delta \mathcal{L}} \rangle_i = \int e^{\lambda \Delta \mathcal{L} + \mathcal{L}_i} \mathcal{D}u \mathcal{D}b,
\end{align}
where the measure denotes an integral over all possible bound and unbound times in the interval $T$. The range of the integral can be split over distinct regions with a particular number of binding events $n$ in time $T$ i.e., $\sum_{j=0}^n u_j + b_j > T$ but $\sum_{j=0}^{n-1} u_j + b_j < T$, where $u_j$ and $b_j$ denote the unbound and bound time at the $j$th binding event. Conditioning over $n$ using the Heaviside function, we have
\begin{align}
M(\lambda) = \sum_{n=1}^{\infty} \int e^{\lambda \Delta \mathcal{L} + \mathcal{L}_i} \left\{\Theta\left(\sum_{j=0}^n u_j + b_j - T\right) - \Theta\left(\sum_{j=0}^{n-1} u_j + b_j - T \right) \right\}  \prod_{j=1}^n du_j db_j
\end{align}
Note that $\mathcal{L}_i$ is a sum of $2n$ independent contributions: $\mathcal{L}_i = \sum_{j=1}^n p_i(u_j) + \sum_{j=1}^n q_i(b_j)$, where $p_i \equiv \log P_i$ and $q_i \equiv \log Q_i$. The Heaviside function $\Theta$ can be replaced by its integral form
\begin{align}
\Theta(x) = \int_{-i\infty}^{i\infty} \frac{d\sigma}{2\pi i \sigma} e^{\sigma x},
\end{align}
where the pole at $\sigma = 0$ should be taken to be in the left half of the complex plane. Applying this expression, we have
\begin{align}
\Theta\left(\sum_j^n u_j + b_j - T\right)  = \int_{-i\infty}^{i\infty} \frac{d\sigma}{2\pi i \sigma} e^{-\sigma T} \prod_j^n e^{\sigma u_j+\sigma b_j}.
\end{align}
Using the same representation for the other $\Theta$ function, we have 
\begin{align}
M(\lambda) = \sum_{n=1}^{\infty} \int_{-i\infty}^{i\infty} \frac{d\sigma}{2\pi i \sigma} e^{-\sigma T} \int \left\{ \prod_{j=1}^n e^{\lambda \Delta p(u_j) + p_i(u_j) + \lambda \Delta q(b_j) + q_i(b_j)} \right\} \left\{ \prod_{j=1}^n e^{\sigma u_j+\sigma b_j} - \prod_{j=1}^{n-1} e^{\sigma u_j+\sigma b_j}\right\}  \prod_{j=1}^n du_j db_j,
\end{align}
where $\Delta p(u_j) = p_0(u_j) - p_1(u_j)$ and similarly for $\Delta q$. Simplifying, we have
\begin{align}
M(\lambda) = \sum_{n=1}^{\infty} \int_{-i\infty}^{i\infty} \frac{d\sigma}{2\pi i \sigma} e^{-\sigma T} \left\{ f(\sigma,\lambda)^n - f(0,\lambda)f(\sigma,\lambda)^{n-1} \right\},
\end{align}
where we have defined 
\begin{align}
f(\sigma,\lambda) = \int_0^{\infty} e^{\lambda \Delta p(u) + p_i(u) + \sigma u} du  \int_0^{\infty}  e^{\lambda \Delta q(b) + q_i(b) + \sigma b} db.
\end{align}
Summing over $n$, we get
\begin{align}\label{eq:Mlambda}
M(\lambda) = \int_{-i\infty}^{i\infty} \frac{d\sigma}{2\pi i \sigma} e^{-\sigma T} \frac{f(\sigma,\lambda) - f(0,\lambda)}{1 - f(\sigma,\lambda)}.
\end{align}
From Cauchy's Residue Theorem, for large $T$, this integral is dominated by the residue from the smallest, positive pole $\sigma$ such that $f(\sigma,\lambda) = 1$, which we call $\sigma_s(\lambda)$. Since we are interested in the cumulants of $\Delta \mathcal{L}$, which are generated by the Taylor series of $\log M(\lambda)$ as $\lambda \to 0$, it is sufficient to consider only small $\lambda$. To see that $\sigma_s$ exists and is positive, we expand $f(\sigma,\lambda)$ around $f(0,0)$ up to first order in a Taylor series
\begin{align}
f(\sigma, \lambda) = f(0,0) + \sigma \frac{\partial f}{\partial \sigma} + \lambda \frac{\partial f}{\partial \lambda} + \dots  .
\end{align}
where all derivatives here and below are evaluated at $(\sigma,\lambda) = (0,0)$. Since $f(0,0) = 1$ and $f(\sigma_s(\lambda), \lambda) = 1$, we have 
\begin{align}
\sigma_s(\lambda) = -\lambda \frac{\partial f/\partial \lambda}{\partial f/\partial \sigma} + O(\lambda^2).
\end{align}
To keep the moment-generating function well-defined, for $i = 0$, we take $\lambda \to 0^{-}$ and for $i=1$, we take $\lambda \to 0^{+}$, which together ensure that for both cases $\sigma_s$ is a positive pole of the integrand in \eqref{eq:Mlambda}. We circumvent the problem of calculating $\sigma_s$ for each $P_i$ and $Q_i$ by noticing that for large $T$, the cumulant generating function $\log M$ can be simply written as
\begin{align}
\log M(\lambda) = -T \sigma_s(\lambda) + o(T). 
\end{align}
The $m$th-order cumulants are then obtained by taking the $m$th-order derivatives of $\sigma_s(\lambda)$ at $\lambda = 0$. Here, we will derive the expressions for the mean and the variance; higher-order cumulants can be obtained by taking further derivatives. As noted above, $f(\sigma_s(\lambda),\lambda) = 1$. Differentiating both sides by $\lambda$, we obtain
\begin{align}\label{eq:sigmasder}
\frac{d \sigma_s}{d\lambda} \frac{\partial f}{\partial \sigma} + \frac{\partial f}{\partial \lambda} = 0
\end{align}
From here, we get
\begin{align}
\langle \Delta \mathcal{L} \rangle_i = -T\frac{d\sigma_s}{d\lambda} = T \frac{\partial f/\partial \lambda}{\partial f/\partial \sigma}
\end{align}
Taking the partial derivatives of $f$ and evaluating at $(0,0)$ gives:
\begin{align}
\frac{\langle \Delta \mathcal{L} \rangle_i}{\langle n \rangle_i} &= \left\langle \mathcal{P}  + \mathcal{Q} \right\rangle_i,  \label{eq:mean}
\end{align}
where we have defined $\mathcal{P} = \log \frac{P_0}{P_1}, \mathcal{Q} = \log \frac{Q_0}{Q_1}$ and $\langle n \rangle_i = \frac{T}{\langle t_u + t_b \rangle_i}$ is the total number of binding events on the receptor in time $T$. 
For instance, at (0,0), we have $\partial f/\partial \lambda = \left\langle \log \frac{P_0}{P_1} + \log \frac{Q_0}{Q_1} \right\rangle_i$ and $\partial f/\partial \sigma = \langle u + b \rangle_i$. The variance can similarly be obtained by applying another derivative w.r.t $\lambda$ on \eqref{eq:sigmasder}. Finally, we get
\begin{align}
\langle  \delta^2 \Delta \mathcal{L} \rangle_i = -T\frac{d^2\sigma_s}{d\lambda^2} = T \frac{\left(\frac{d \sigma_s}{d\lambda}  \right)^2 \frac{\partial^2 f}{\partial \sigma^2} + 2\frac{d \sigma_s}{d\lambda} \frac{\partial^2 f}{\partial \sigma \partial \lambda} + \frac{\partial^2 f}{\partial \lambda^2} }{\frac{\partial f}{\partial \sigma}}.
\end{align}
The expression above can be evaluated as in the examples above to obtain:
\begin{align}
\frac{\langle \delta^2 \Delta \mathcal{L} \rangle_i}{\langle n \rangle_i} = \left\langle \mathcal{P}  +   \mathcal{Q} \right\rangle_i^2 \frac{\langle (t_u + t_b)^2 \rangle_i}{\langle t_u + t_b \rangle_i^2} & - 2\left\langle \mathcal{P}  +  \mathcal{Q} \right\rangle_i\frac{\langle (t_u + t_b)(\mathcal{P} + \mathcal{Q})\rangle_i}{\langle t_u + t_b \rangle_i} \label{eq:var} +  \langle (\mathcal{P} + \mathcal{Q})^2\rangle_i.
\end{align}

\begin{figure}[t!]
\includegraphics[width=0.8\linewidth]{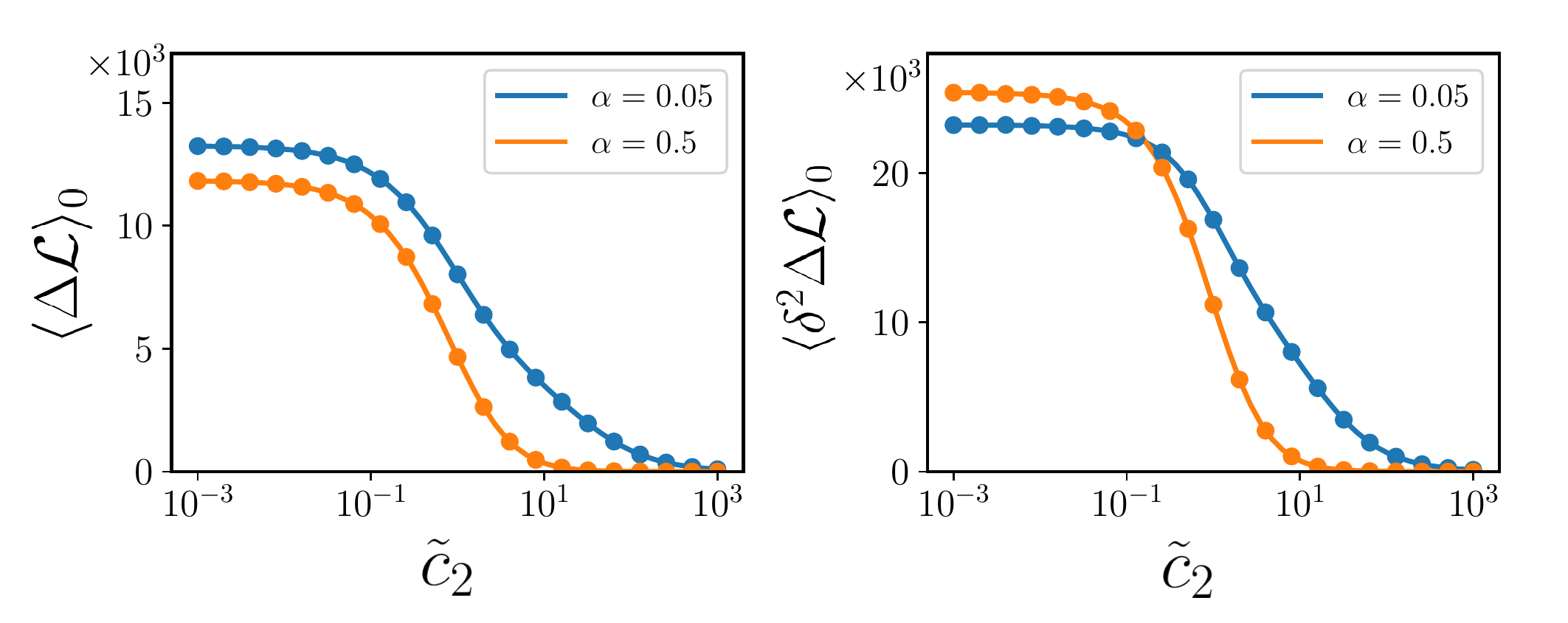}
\caption{The values from expressions for the mean (\eqref{eq:mean}, solid lines, left panel) and variance (\eqref{eq:var}, solid lines, right panel) of the log-likelihood differences, $\Delta \mathcal{L}$, align with those obtained from numerical simulations (circles). Parameters are the same as those in Figure \ref{fig:discriminability}.\label{fig:disc_formula}}
\end{figure}

\section{Optimizing the proofreading scheme}
The variance of $\hat{x}_A$ is
\begin{align}
\delta x_A^2 = \frac{f_A  ( 1- f_A)}{n_A (e^{-\alpha r' \tau_A} - e^{-r' \tau_A})^2} ,
\end{align}
where $f_A = xe^{-r \tau_A} + \left(1-x \right)e^{-r'\tau_A}$. We define $y_A = e^{-r'\tau_A}$ and write the above equation as 
\begin{align}
\delta x_A^2 \times n_A  = \frac{y_A(1-y_A)}{(y_A^{\alpha} - y_A)^2} + x \frac{1-2y_A}{y_A^{\alpha} - y_A} -x^2 ,
\end{align}
To make the RHS of the above expression tractable for optimization, we observe that we are imposing a cutoff that discriminates between samples drawn from exponential distributions of means with ratio $\alpha^{-1}$. As $\alpha \to 0$, the ratio gets larger and we expect the threshold to be placed much greater than mean bound time of non-cognate binding events, $\sim r'^{-1}$. Accordingly, we propose an ansatz that $y_A^{\alpha} = O(1)$ and $y_A \ll 1$ as $\alpha \to 0$. Using the ansatz, we simplify the equation above to get
\begin{align}
\delta x_A^2 \times  n_A  \approx y_A^{1-2\alpha} + xy_A^{-\alpha}.
\end{align}
Optimizing the right hand side of the above equation w.r.t $y_A$ by taking the derivative in $y_A$ and equating to zero, we get
\begin{align}\label{eq:tauAopt}
r'\tau_A^* \approx (1-\alpha)^{-1} \log{\left(\frac{1-2\alpha}{\alpha x} \right)},
\end{align}
which is reproduced in the main text and used in Figure \ref{fig:networkerror}. We verify then that as $\alpha \to 0$, $e^{-r' \tau_A^*} \ll 1$ and since $\alpha^{\alpha} \to 1$, we have $e^{-\alpha r' \tau_A^*}  = O(1)$. A similar expression can be derived for $\tau_B^*$ with $x$ replaced by $1-x$ in \eqref{eq:tauAopt}.

\section{Nonspecific ligand pool}

With total concentration given by $c_{tot}=c_{1}+c_{2}+c_{0}$, the
probability of a sequence of bound and unbound times $\{u_{i}^{R},b_{i}^{R}\}$
can be written as,

\begin{eqnarray}
P(\{u_{i}^{R},b_{i}^{R}\}) & ={\displaystyle \prod_{R=A,B}\:}{\displaystyle \prod_{i=1}^{n_{R}}}e^{-k_{R}c_{tot}u_{i}^{R}}\Big[k_{R}c_{1}r_{R,1}e^{-r_{R,1}b_{i}^{R}}+k_{R}c_{2}r_{R,2}e^{-r_{R,2}b_{i}^{R}}\nonumber \\
 & +k_{R}c_{0}r_{R,0}e^{-r_{R,0}b_{i}^{R}}\Big],
\end{eqnarray}
with index $R$ running over the receptors $A$ and $B$, and index
$i$ running over the binding events $n_{R}$ in a fixed interval
$T$. Now, log-likelihood $\mathcal{L}(x,y,c_{tot})$ as a function
of the fractions of the cognate ligands, $x$ and $y$, and the total
concentration, $c_{tot}$, is given by,
\begin{eqnarray}
\mathcal{L} & = & {\displaystyle \sum_{R=A,B}\Big\{-k_{R}c_{tot}T_{u}^{R}+n_{R}\log k_{R}r_{R,0}c_{tot}-r_{R,0}T_{b}^{R}}\nonumber \\
 &  & +{\displaystyle \sum_{i=1}^{n_{R}}\log\big(x\lambda_{R,1}e^{(1-\lambda_{R,1})r_{R,0}b_{i}^{R}}+y\lambda_{R,2}e^{(1-\lambda_{R,2})r_{R,0}b_{i}^{R}}+1-x-y\big)}\Big\},
\end{eqnarray}
where $R$ runs over the receptors, $i$ runs over the binding events
(a total of $n_{R}$) of each receptor, $b_{i}^{R}$ denotes the bound
time of the $i^{th}$ binding event of receptor $R$, $T_{u}^{R}$
and $T_{b}^{R}$ denote the total unbound and bound time respectively
of receptor $R$. We also define $\lambda_{R,j}\equiv\frac{r_{R,j}}{r_{R,0}}$
as the ratio of the unbinding rate of ligand $j$ to the unbinding
rate of the nonspecific ligand from receptor $R$. By setting $\frac{\partial\mathcal{L}(\bm{\theta})}{\partial\bm{\theta}}\rvert_{\bm{\theta}^{*}}=0$,
we obtain the ML estimates of $\bm{{\theta}}=[x,y,c_{tot}]$. The
ML estimate of total concentration, $c_{tot}^{*}$, is
\begin{equation}
c_{tot}^{*}=\dfrac{n_{A}+n_{B}}{k_{A}T_{u}^{A}+k_{B}T_{u}^{B}}.
\end{equation}
The ML esimates of the fractions of cognate ligands, $x^{*}$ and
$y^{*}$, satisfy the following equations.
\begin{eqnarray}
{\displaystyle \sum_{R=A,B}}\:{\displaystyle \sum_{i=1}^{n_{R}}\dfrac{\lambda_{R,1}e^{(1-\lambda_{R,1})r_{R,0}b_{i}^{R}}-1}{x^{*}\lambda_{R,1}e^{(1-\lambda_{R,1})r_{R,0}b_{i}^{R}}+y^{*}\lambda_{R,2}e^{(1-\lambda_{R,2})r_{R,0}b_{i}^{R}}+1-x^{*}-y^{*}}}=0,\\
{\displaystyle \sum_{R=A,B}}\:{\displaystyle \sum_{i=1}^{n_{R}}\dfrac{\lambda_{R,2}e^{(1-\lambda_{R,2})r_{R,0}b_{i}^{R}}-1}{x^{*}\lambda_{R,1}e^{(1-\lambda_{R,1})r_{R,0}b_{i}^{R}}+y^{*}\lambda_{R,2}e^{(1-\lambda_{R,2})r_{R,0}b_{i}^{R}}+1-x^{*}-y^{*}}}=0.
\end{eqnarray}
Next, by taking expectation over the distribution of the sequence
of bound and unbound times, we obtain $\langle\frac{\partial\mathcal{L}}{\partial\bm{\theta}}\rangle$,
which informs us whether the ML estimates are unbiased, and $\langle\frac{\partial^{2}\mathcal{L}}{\partial\bm{\theta}^{2}}\rangle$,
which help us obtain the Fisher Information matrix, $I(\bm{\theta})$.
We note that this probability $P$ factorizes and can be written as,
\begin{equation}
P(\{u_{i}^{R},b_{i}^{R}\})=\prod_{R=A,B}\:\prod_{i=1}^{n_{R}}p(u_{i}^{R})p(b_{i}^{R}).
\end{equation}
It is trivial to show that $\langle\frac{\partial\mathcal{L}}{\partial\bm{\theta}}\rangle=0$
and hence the ML estimates $\bm{\theta^{*}}$ are unbiased. Now, setting
$u=r_{R,0}b_{i}^{R}$, we have,
\begin{eqnarray}
\Big\langle\dfrac{\partial^{2}\mathcal{L}}{\partial x^{2}}\Big\rangle=-\sum_{R=A,B}\langle n_{R}\rangle\int_{0}^{+\infty}due^{-u}\dfrac{(\lambda_{R,1}e^{(1-\lambda_{R,1})u}-1)^{2}}{x\lambda_{R,1}e^{(1-\lambda_{R,1})u}+y\lambda_{R,2}e^{(1-\lambda_{R,2})u}+1-x-y}\label{eq:dL2dx2},\\
\Big\langle\dfrac{\partial^{2}\mathcal{L}}{\partial y^{2}}\Big\rangle=-\sum_{R=A,B}\langle n_{R}\rangle\int_{0}^{+\infty}due^{-u}\dfrac{(\lambda_{R,2}e^{(1-\lambda_{R,2})u}-1)^{2}}{x\lambda_{R,1}e^{(1-\lambda_{R,1})u}+y\lambda_{R,2}e^{(1-\lambda_{R,2})u}+1-x-y},\\
\Big\langle\dfrac{\partial^{2}\mathcal{L}}{\partial x\partial y}\Big\rangle=-\sum_{R=A,B}\langle n_{R}\rangle\int_{0}^{+\infty}due^{-u}\dfrac{(\lambda_{R,1}e^{(1-\lambda_{R,1})u}-1)(\lambda_{R,2}e^{(1-\lambda_{R,2})u}-1)}{x\lambda_{R,1}e^{(1-\lambda_{R,1})u}+y\lambda_{R,2}e^{(1-\lambda_{R,2})u}+1-x-y}.
\end{eqnarray}
We also have 
\begin{eqnarray}
 &  & \Big\langle\dfrac{\partial^{2}\mathcal{L}}{\partial c_{tot}^{2}}\Big\rangle=-\frac{\langle n_{A}\rangle+\langle n_{B}\rangle}{c_{tot}^{2}},\\
 &  & \Big\langle\dfrac{\partial^{2}\mathcal{L}}{\partial x\partial c_{tot}}\Big\rangle=0,\\
 &  & \Big\langle\dfrac{\partial^{2}\mathcal{L}}{\partial y\partial c_{tot}}\Big\rangle=0\label{eq:dLdxdLdc}.
\end{eqnarray}
Using Eq. \ref{eq:dL2dx2}-\ref{eq:dLdxdLdc}, the Fisher Information
matrix $I(\bm{\theta})$ can be obtained as $-\langle\frac{\partial^{2}\mathcal{L}}{\partial\bm{\theta}^{2}}\rangle$.
The Cram\'{e}r-Rao bound, in the limit of a large number of binding events,
which can be ensured by choosing $T$ to be significantly longer than
a typical binding/unbinding event, ensures that
\begin{eqnarray}
\Sigma_{\bm{\theta}}=I(\bm{\theta})^{-1}.
\end{eqnarray}
We are interested in the log concentrations, $l_{i}=\log c_{i}$.
What is an appropriate cost-function in this case? With $\bm{\theta'}=[l_{1},l_{2}]$,
and $\bm{l_{all}}=[l_{1},l_{2},l_{0}]=[\bm{\theta'},l_{0}]$, we obtain
the covariance matrix $\Sigma_{\bm{l_{all}}}$ as 
\begin{eqnarray}
\Sigma_{\bm{l_{all}}}=J_{all}\Sigma_{\bm{\theta}}J_{all}^{T},
\end{eqnarray}
where $J_{all}$ is the Jacobian given by
\begin{eqnarray}
\begin{pmatrix}x^{-1} & 0 & c_{tot}^{-1}\\
0 & y^{-1} & c_{tot}^{-1}\\
-z^{-1} & -z^{-1} & c_{tot}^{-1}
\end{pmatrix}.
\end{eqnarray}
This describes an ellipsoid in $\bm{l_{all}}$ space. We are interested
only in $\bm{\theta'}$. So we obtain the marginal distribution of
$l_{1}$ and $l_{2}$ and obtain the covariance matrix, $\Sigma_{\bm{\theta'}}=J\Sigma_{\bm{\theta}}J^{T}$,
and the cost function $\Omega_{NS}$ as the area of the ellipse centered
at $(l_{1},l_{2})$ as
\begin{eqnarray}
\Omega_{NS}=\det(\Sigma_{\bm{\theta'}})=\det(J\Sigma_{\bm{\theta}}J^{T}),
\end{eqnarray}
where the Jacobian $J$ given by 
\begin{eqnarray}
\begin{pmatrix}x^{-1} & 0 & c_{tot}^{-1}\\
0 & y^{-1} & c_{tot}^{-1}
\end{pmatrix}.
\end{eqnarray}

\end{document}